\documentclass{mem}
\usepackage{natbib}
\usepackage{txfonts}
\usepackage{balance}
\usepackage{graphicx}
\usepackage[a4paper,breaklinks,dvipdfm]{hyperref}
\idline{75}{282}
\begin{document}
\def\teff{$T\rm_{eff }$}
\def\kms{$\mathrm {km s}^{-1}$}

\title{Detecting intermediate mass black holes in globular clusters with machine learning}

   \subtitle{}

\author{
M. \,Pasquato\inst{1} 
          }
          
\institute{Department of Astronomy \& Center for Galaxy Evolution ResearchYonsei University, 120-749, Seoul, Republic of Korea \email{mario.pasquato@galaxy.yonsei.ac.kr}} 

\authorrunning{Pasquato}

\titlerunning{Detecting IMBHs in GCs with machine learning}

\abstract{Mergers of stellar-mass black holes were recently observed in the gravitational wave window opened by LIGO. This puts the spotlight on dense stellar systems and their ability to create intermediate-mass black holes (IMBHs) through repeated merging. Unfortunately, attempts at direct and indirect IMBH detection in star clusters in the nearby universe have proven inconclusive as of now. Indirect detection methods attempt to constrain IMBHs through their effect on star cluster photometric and kinematic observables. They are usually based on looking for a specific, physically motivated signature. While this approach is justified, it may be suboptimal in its usage of the available data. Here I present a new indirect detection method, based on machine learning, that is unaffected by these restrictions. I reduce the scientific question whether a star cluster hosts an IMBH to a classification problem in the machine learning framework. I present preliminary results to illustrate how machine learning models are trained on simulated dataset and measure their performance on previously unseen, simulated data.
}
\maketitle{}

\section{Introduction}
The recent detections of black-hole mergers by LIGO \citep[][]{2016PhRvL.116f1102A} showed that stellar-mass black holes in the local universe indeed do merge, giving rise to heavier objects. This finding supports the intermediate-mass black hole (IMBH) formation scenario based on repeated mergers in dense stellar systems \citep[][]{2002MNRAS.330..232C} such as massive star clusters. IMBHs are thus expected to be present in at least some clusters in the Milky Way, and should be detectable either directly (through radio or X-ray emission) or indirectly (by their effect on cluster dynamics). Until now, however, both the direct and indirect approach were inconclusive, with no undisputed detection \citep[see e.g.][]{2011A&A...533A..36L, 2013ApJ...769..107L}. Direct detection in old stellar systems such as globular clusters is intrinsically difficult due to the lack of gas in that environment, while indirect detection requires care to optimise the usage of the available kinematic and photometric data. In this respect, indirect detection methods are usually based on looking for a specific, physically motivated signature, so they may potentially throw away a large part of the information contained in the data. In this paper I present an indirect detection method based on machine learning. I create a synthetic sample of star clusters by running direct N-body simulations. A fraction of the clusters contain an IMBH. I then prepare mock observations from simulation snapshots and measure cluster observables. Each snapshot is thus mapped into a point in an N-dimensional \emph{feature space}, where machine learning algorithms are applied to classify clusters into IMBH hosts or non-hosts. The classifiers are then used to predict the classification of previously unseen simulated data. 

\section{Simulations}
I run a set of N-body simulations using the direct summation code NBODY6 \citep[][]{1999PASP..111.1333A}. The initial conditions for all simulations are \cite{1966AJ.....71...64K} models with central dimensionless potential in the $2 - 8$ range, with no primordial binaries and equal mass stars, except for the IMBH (when present). The simulations were evolved for $1000$ N-body units \citep[][]{1986LNP...267..233H}, corresponding to about three half-mass relaxation times. I varied the mass of the IMBH in the $50 - 250$ range in units of the mass of a cluster star. The simulated clusters evolved in isolation (i.e. without tidal interaction with the host galaxy) and no stellar evolution was considered. Some simulations share initial conditions but were initialised with a different random seed. The simulations are listed in Tab.~\ref{sims}.

\begin{table}
\caption{Simulation set. Each simulation is identified by a string summarizing its initial conditions (Col.~1). The simulation 16kBH50W02 for example contains $16000$ stars, an IMBH with mass $50 M_{*}$ where $M_{*}$ is the mass of a cluster star in the simulation, and was initialised as a King model with central dimensionless potential $W_0 = 2$. When the identifier ends in \emph{s} followed by a number it is a rerun with a different random seed. The number of stars in each simulation is listed in Col.~2, the black hole mass in Col.~3, and $W_0$ in Col.~4.}
\label{sims}
\begin{center}
\begin{tabular}{llrl}
\hline
\\
Simulation & \# stars $/{{10}^3}$ & $M_{BH}/M_{*}$ & $W_0$\\
\hline
\\
16kBH50W02 &$16$ & $50.0$& $2$\\
16kBH50W04 &$16$ & $50.0$& $4$\\
16kBH50W06 &$16$ & $50.0$& $6$\\
16kBH50W08 &$16$ & $50.0$& $8$\\
16kBH100W02 &$16$ & $100.0$& $2$\\
16kBH100W04 &$16$ & $100.0$& $4$\\
16kBH100W06 &$16$ & $100.0$& $6$\\
16kBH100W08 &$16$ & $100.0$& $8$\\
16kBH150W02 &$16$ & $150.0$& $2$\\
16kBH150W04 &$16$ & $150.0$& $4$\\
16kBH150W06 &$16$ & $150.0$& $6$\\
16kBH200W02 &$16$ & $200.0$& $2$\\
16kBH200W04 &$16$ & $200.0$& $4$\\
16kBH200W06 &$16$ & $200.0$& $6$\\
16kBH250W02 &$16$ & $250.0$& $2$\\
\\
16kNOBHW02 &$16$ & $0.0$& $2$\\
16kNOBHW02s2 &$16$ & $0.0$& $2$\\
16kNOBHW02s3 &$16$ & $0.0$& $2$\\
16kNOBHW02s345 &$16$ & $0.0$& $2$\\
16kNOBHW02s4 &$16$ & $0.0$& $2$\\
16kNOBHW03 &$16$ & $0.0$& $3$\\
16kNOBHW03s341 &$16$ & $0.0$& $3$\\
16kNOBHW04 &$16$ & $0.0$& $4$\\
16kNOBHW04s374 &$16$ & $0.0$& $4$\\
16kNOBHW04s60 &$16$ & $0.0$& $4$\\
16kNOBHW04s70 &$16$ & $0.0$& $4$\\
16kNOBHW04s80 &$16$ & $0.0$& $4$\\
16kNOBHW06 &$16$ & $0.0$& $6$\\
16kNOBHW06s60 &$16$ & $0.0$& $6$\\
16kNOBHW06s70 &$16$ & $0.0$& $6$\\
16kNOBHW06s80 &$16$ & $0.0$& $6$\\
16kNOBHW08 &$16$ & $0.0$& $8$\\
16kNOBHW08s60 &$16$ & $0.0$& $8$\\
16kNOBHW08s70 &$16$ & $0.0$& $8$\\
16kNOBHW08s80 &$16$ & $0.0$& $8$\\
\\
\hline
\end{tabular}
\end{center}
\end{table}

\section{Mock observations, feature space, dimensionality reduction, and learning}
I extracted $700$ snapshots from the simulations ($20$ snapshots spaced by $10$ N-body units - about four crossing times - for each simulation). The positions and velocities of a randomly selected fraction of the stars in each snapshot (to simulate observational incompleteness) were converted to projected values (radial distance on the plane of the sky from the cluster center and velocity along the line of sight), obtaining a two-dimensional plot in the radius-velocity plane for each snapshot. The plane was then overlaid with a square NxN grid, resulting in 2D bins within which the number of stars was counted and normalised to the $[0, 1]$ range. This translated every snapshot into an image, i.e. an array of $N^2$ numeric values comprised between $0$ and $1$. The feature space is thus $N^2$ dimensional. The effect of different values of $N$ was explored, but in any case the large dimensionality of the feature space called for dimensionality reduction, which was carried out with principal component analysis. Only the first $10$ principal components were retained. On this dimension-reduced feature space I trained plain C5.0 trees \citep[][]{1993cpml.book.....Q} using the R library C50.

\section{Validation}
I measured the accuracy of classification by using five-fold cross-validation. In this approach the dataset is randomly partitioned into five subsets, each sharing the same number of records. Training algorithms are then applied to four of the five slices and the trained model is used to predict the classification label of data in the fifth slice. This is repeated five times rotating the slices. The resulting predictions are compared to the actual label of the data (i.e. to whether a snapshot contained an IMBH or not) and a rate of misclassification is computed as follows:
\begin{equation}
\label{miscl}
D = \frac{\mathrm{\# misclassifications}}{\mathrm{\# snapshots}}
\end{equation}

The cross validation procedure was repeated $10$ times with different random seeds to estimate the standard deviation of the distribution of $D$. 
\section{Results and conclusions}

The misclassification rates $D$ obtained using cross-validation range from few percent to over $20\%$ depending on the parameters chosen. In particular I observed a dependence on
\begin{itemize}
\item the field of view of the mock observations;
\item the number of 2D bins in the mock observations;
\item the completeness of the mock observations;
\item the number of principal components of feature space included in the analysis;
\end{itemize}

\begin{figure}
\resizebox{0.99\columnwidth}{!}{
\includegraphics{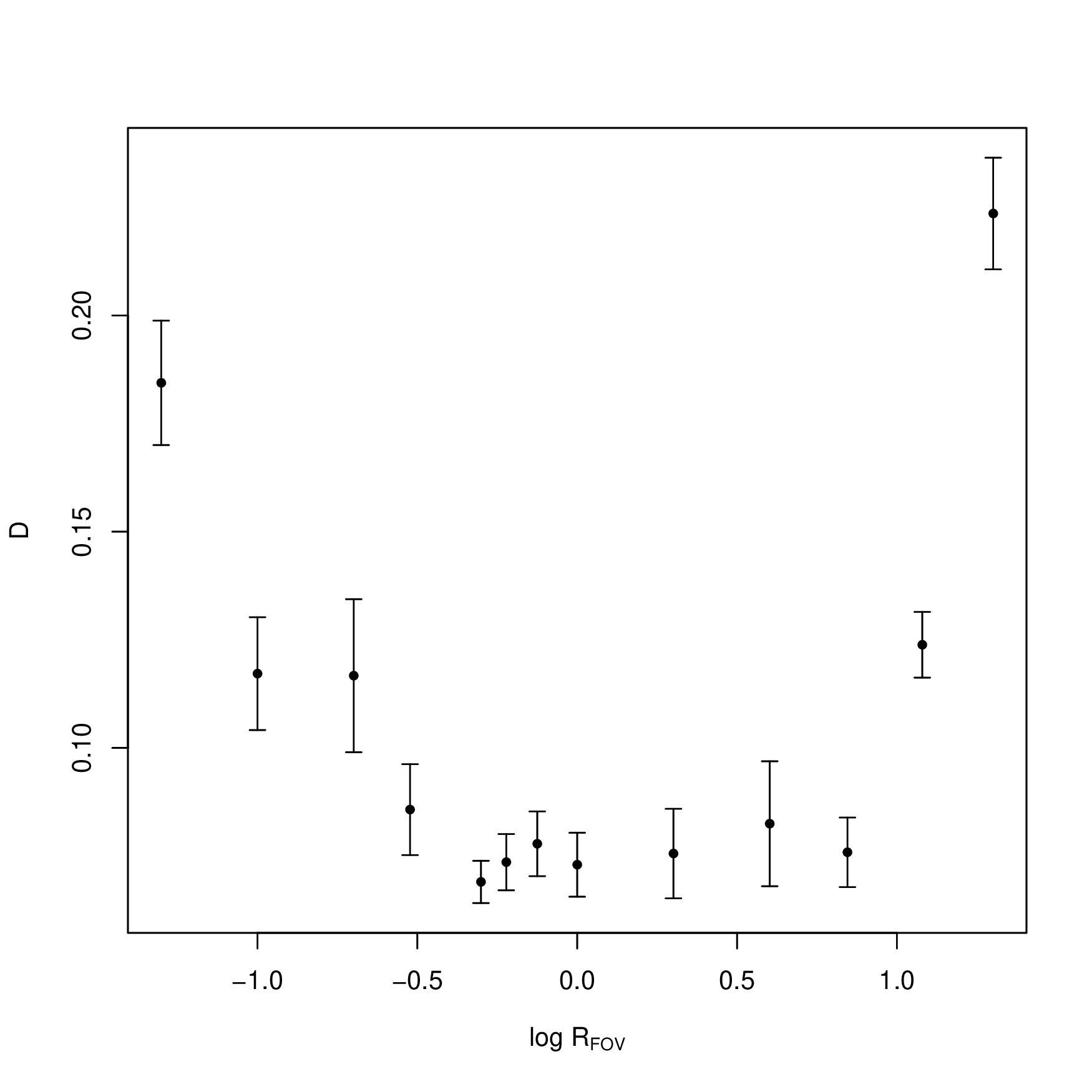}}
\caption{\footnotesize
Fraction of misclassified snapshots as a function of the size of the field of view (in units of the projected half-mass radius of the simulated cluster) for a fixed number of 2D bins ($30 \times 30$). Increasing the field of view improves the accuracy for small fields of view because more relevant information is accessible, but degrades the accuracy for large fields of view because it uses up 2D bins in the external regions of the cluster, thus reducing the resolution in the center due to the fixed number of bins.\label{FOV}
}
\end{figure}

In Fig.~\ref{FOV} I plot the misclassification rate $D$ as a function of the size of the field of view (in units of the projected half-mass radius of the simulated cluster) for a fixed number of 2D bins ($30 \times 30$) and principal components used ($10$). 

\begin{figure}
\resizebox{0.99\columnwidth}{!}{
\includegraphics{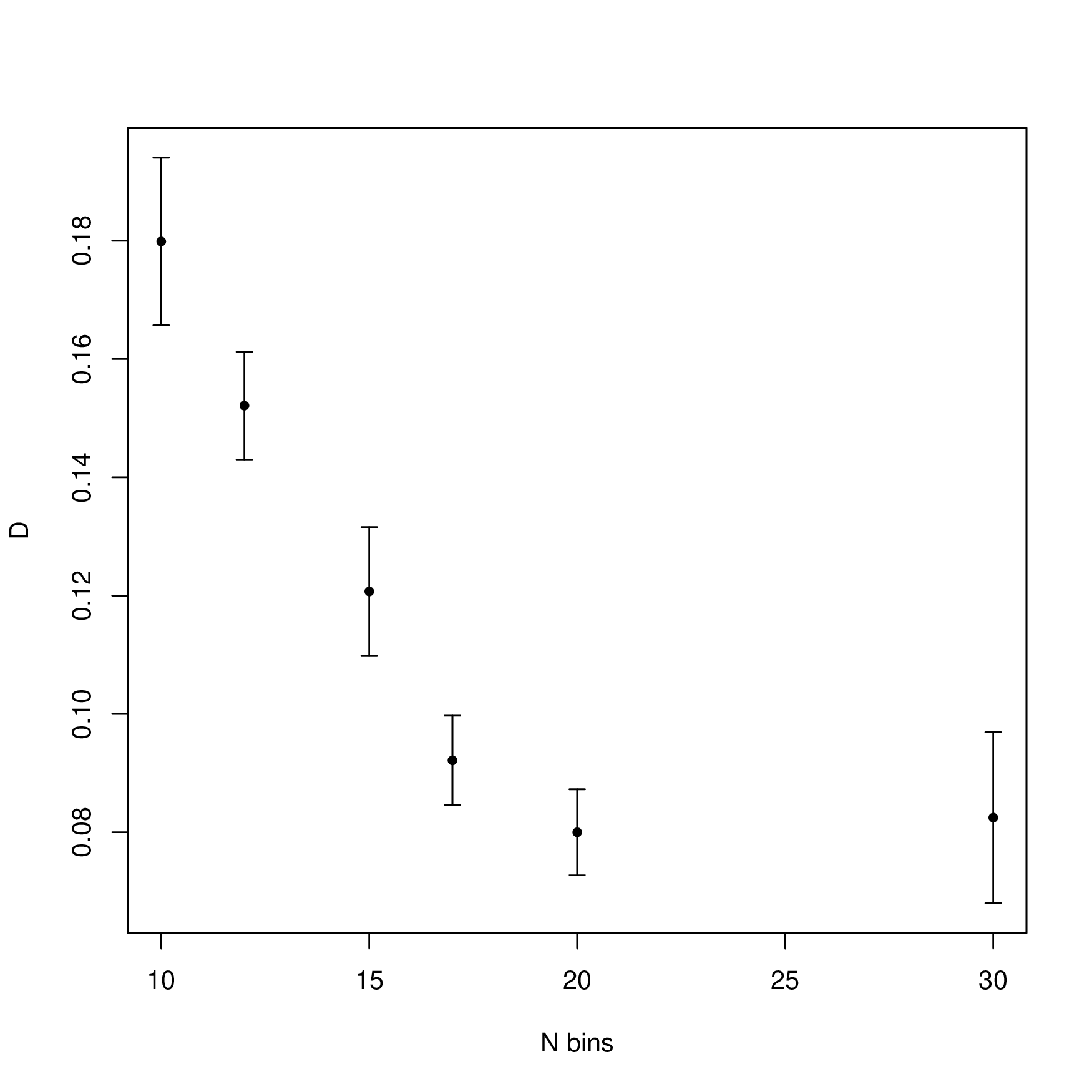}}
\caption{
\footnotesize
Fraction of misclassified snapshots as a function of the number of 2D bins for a fixed field of view. Increasing the number of bins monotonically improves the accuracy, but a larger number of bins slows down the calculations with increasingly reduced effect on the accuracy. \label{Np}
}
\end{figure}

In Fig.~\ref{Np} the number of 2D bins is varied for a fixed field of view equal to four times the half-mass radius and $10$ principal components. In both cases the completeness was set to $1$.

In this simpified set-up an out-of-the-box algorithm such as C$5.0$ achieves a misclassification rate of some percent in the most favourable cases. This is encouraging, suggesting that further development of machine-learning based indirect detection may be a promising way to spot IMBHs in real observational data.

\begin{acknowledgements}
I acknowledge support from Mid-career Researcher Program (No. 2015-008049) through the National Research Foundation (NRF) of Korea.
\end{acknowledgements}

\bibliographystyle{aa}

\begin{thebibliography}{}
\bibitem[Aarseth(1999)]{1999PASP..111.1333A} Aarseth, S.~J.\ 1999, \pasp, 111, 1333 
\bibitem[Abbott et al.(2016)]{2016PhRvL.116f1102A} Abbott, B.~P., Abbott, R., Abbott, T.~D., et al.\ 2016, Physical Review Letters, 116, 061102  
\bibitem[King(1966)]{1966AJ.....71...64K} King, I.~R.\ 1966, \aj, 71, 64 
\bibitem[Heggie \& Mathieu(1986)]{1986LNP...267..233H} Heggie, D.~C., \& Mathieu, R.~D.\ 1986, The Use of Supercomputers in Stellar Dynamics, 267, 233
\bibitem[Lanzoni et al.(2013)]{2013ApJ...769..107L} Lanzoni, B., Mucciarelli, A., Origlia, L., et al.\ 2013, \apj, 769, 107 
\bibitem[L{\"u}tzgendorf et al.(2011)]{2011A&A...533A..36L} L{\"u}tzgendorf, N., Kissler-Patig, M., Noyola, E., et al.\ 2011, \aap, 533, A36 
\bibitem[Miller \& Hamilton(2002)]{2002MNRAS.330..232C} Miller, M.~C., \& Hamilton, D.~P.\ 2002, \mnras, 330, 232 
\bibitem[Pasquato \& Chung(2016)]{2016A&A...589A..95P} Pasquato, M., \& Chung, C.\ 2016, \aap, 589, A95  
\bibitem[Quinlan(1993)]{1993cpml.book.....Q} Quinlan, J.~R.\ 1993, The Morgan Kaufmann Series in Machine Learning, San Mateo, CA: Morgan Kaufmann, |c1993,  

\end{thebibliography}

\end{document}